\documentstyle[11pt,epsfig,epsf]{article}

\newlength{\bredde}
\def\slash#1{\settowidth{\bredde}{$#1$}\ifmmode\,\raisebox{.15ex}{/}
\hspace*{-\bredde} #1\else$\,\raisebox{.15ex}{/}\hspace*{-\bredde} #1$\fi}
\newcommand{\be}{\begin{eqnarray}}
\newcommand{\ee}{\end{eqnarray}}
\newcommand{\beq}{\begin{equation}}
\newcommand{\eeq}{\end{equation}}
\newcommand{\ba}{\begin{array}{ccc}}
\newcommand{\ea}{\end{array}}
\newcommand{\nn}{\nonumber}
\newcommand{\noi}{\vspace{12pt}\noindent}
\def\al{\alpha}
\def\zc{{z^\ast}}
\def\muc{\mu_{chem}}
\def\tr{{\mbox{Tr}}}
\def\re{{\Re\mbox{e}}}
\def\im{{\Im\mbox{m}}}
\begin{document}
\topmargin -1.4cm
\oddsidemargin -0.8cm
\evensidemargin -0.8cm
\title{\Large{{\bf 
Microscopic correlations of non-Hermitian Dirac operators in 
three-dimensional QCD}}} 

\vspace{1.5cm}

\author{~\\{\sc G. Akemann}\\~\\Max-Planck-Institut f\"ur Kernphysik\\
Saupfercheckweg 1\\D-69117 Heidelberg\\Germany\\
}
\date{}
\maketitle
\vfill
\begin{abstract}
In the presence of a non-vanishing chemical potential the eigenvalues 
of the Dirac operator become complex. We calculate spectral 
correlation functions of complex eigenvalues using a random matrix 
model approach. Our results apply to non-Hermitian Dirac operators 
in three-dimensional QCD with broken flavor symmetry and in 
four-dimensional QCD in the bulk of the spectrum. The derivation follows 
earlier results of Fyodorov, Khoruzhenko and Sommers for complex spectra 
exploiting the existence of orthogonal polynomials in the complex plane. 
Explicit analytic expressions are given for all microscopic 
$k$-point correlation functions in the presence of an arbitrary even 
number of massive quarks,  both in the limit of strong and weak 
non-Hermiticity. In the latter case the parameter governing the 
non-Hermiticity of the Dirac matrices is identified with the influence 
of the chemical potential.
\end{abstract}
\vfill

\thispagestyle{empty}
\newpage

\setcounter{equation}{0}
\section{Introduction}

Random Matrix Theory (RMT) does not only provide a very useful tool
for spectra of real variables such as energy levels. Also in the case
where the eigenvalues of the underlying Hamiltonian become complex RMT 
describes universal properties in a variety of different physical
models such as localization in superconductors \cite{HN}, 
dissipation and scattering 
in Quantum Chaos \cite{Haa,GHS,Ef,FK} or chiral symmetry 
breaking in Quantum Chromodynamics (QCD) \cite{Steph}. 
It is the latter subject
which has motivated the results presented here. In the presence of a
non-vanishing chemical potential for the quarks the eigenvalues of the 
Dirac operator become complex. The regime with broken chiral (or
flavor) symmetry which we want to describe is essentially non-perturbative
in nature. One of the prominent techniques starting from first
principles is to perform Monte-Carlo simulations on a lattice. However, 
in the presence of complex eigenvalues the simulations run into serious 
problems due to the presence of a complex determinant of the Dirac
operator, which have not yet been overcome to large extent (for a
review see \cite{latt} and references therein).
A better analytical understanding of the correlation functions of
Dirac eigenvalues
in the vicinity of the origin which are very sensitive to the symmetry
breaking would therefore be very useful. The question of having
better control over the lattice simulations in a certain regime is not of
academic nature since the chiral phase transitions is now also
investigated experimentally.

In the past few years RMT has been developed as a powerful analytic
tool to investigate this topic (for a recent review see
\cite{VW}). The reason for its applicability has by now been well
understood on field theoretic grounds, analytic expressions have been
provided for correlation functions at vanishing and non-vanishing
temperature and even the qualitative picture of the QCD phase diagram
has been understood in terms of RMT. However, analytic expressions for 
microscopic correlations in the presence of a chemical potential have
not been derived so far and the present article is meant to partially
close this gap. The only exception is the quenched limit with
zero quark flavors. 
We will recall that the 
quenched correlations follow from \cite{FKS97} for
weakly and from \cite{Gin} for strongly non-Hermitian 
spectra. The nearest neighbor spacing distribution is also known in
both cases, \cite{FKS97} and \cite{GHS} respectively, but not in a 
closed form. It has been already compared to quenched QCD lattice data 
with chemical potential in \cite{MPW} and a cross-over has been
observed. Finally there are results \cite{Janik} 
for the spectral rigidity following semi-classical arguments.

We will focus on the most simple model, the Hermitian matrix model or
Unitary Ensemble, and discuss its generalization to incorporate
complex eigenvalues in the presence of massive quarks. 
This model is relevant for Euclidean QCD in three
dimensions (QCD$_3$) \cite{VZ} and for the correlations in the bulk 
in four dimensions (QCD$_4$) as observed in \cite{HV,GMMW}.
Because of the absence of chiral symmetry
in odd dimensions the corresponding global symmetry is flavor symmetry,
which can be spontaneously broken from $U(2N_f)\to U(N_f)\times
U(N_f)$ for an even number of quark flavors $2N_f$. Recently 
evidence has been provided from lattice data \cite{DHKM} 
for the existence of a non-vanishing condensate in QCD$_3$.
In the following we will assume to be in the broken phase. In
particular we will not address the question of different phases at high 
density from the RMT point of view \cite{Jack}.

The corresponding correlation functions at vanishing chemical
potential have been first derived from RMT for massless \cite{VZ} and massive
flavors \cite{DNII} and proven to be universal \cite{ADMN,DNII}. It
has been observed that they can be related to finite-volume partition
functions \cite{AD}.
Very recently all correlation functions have then  
been derived entirely from the underlying chiral Lagrangian using
supersymmetry \cite{Szabo} and using the replica method \cite{ADDV},
including the corresponding sum-rules \cite{ADDV,DV}. 
These results justify the RMT approach a posteriori and put it on firm
field-theoretic grounds. 

Our approach will be more phenomenological here as it does not yet follow
from a chiral Lagrangian. 
In complete analogy to QCD$_4$ a chemical potential $\muc$ can be
introduced on the level of the QCD$_3$
Lagrangian for the quarks, leading to a shift of the time derivative inside the
Dirac operator $\partial_0\to\partial_0+\muc$. 
This step clearly renders the Dirac eigenvalues to
be complex. 
In the RMT formulation of Euclidean QCD$_4$ at $\muc=0$ the Dirac operator is 
replaced by an off-diagonal block matrix of complex matrices,
$\mbox{offdiag}(W,W^\dagger)$. Its eigenvalues are real and occur in
pairs of opposite sign, reflecting chiral symmetry. 
The insertion of a chemical potential is 
then mimicked by shifting both blocks by 
$W^{(\dagger)}\to
W^{(\dagger)}+i\muc$. After this step, the RMT eigenvalues 
are complex. So far no analytic results for the microscopic
correlation functions have been obtained as a function of $\muc$. 

In order to make some progress we address the 
RMT of QCD$_3$ which is much simpler. 
Here, the Euclidean Dirac operator is replaced by a
random matrix $H$ which is Hermitian. The flavor symmetry breaking
becomes visible after switching on quark masses (for a detailed
discussion of symmetry breaking 
in QCD$_3$ see 
e.g. \cite{Szabo}). In QCD$_3$ these masses 
are purely imaginary and have to occur in pairs of opposite sign
\cite{VZ}, the quark mass matrix reading 
diag$(im_1,\ldots,im_{N_f},-im_1,\ldots,-im_{N_f})$. The full
determinant of the Dirac operator is therefore given by
$\prod_{f=1}^{N_f}\det[H-im_f]\det[H+im_f]$ in the presence of $2N_f$ quark
flavors. As a first attempt we introduce the chemical potential $\muc$ 
in the same way as it was proposed 
in the RMT for QCD$_4$ by simply shifting $H\to H+i\muc$. 
However, this shift is trivial\footnote{Similar to shifting $H$
  by a real constant $c$, $H\to H+c$, which   
leaves the bulk correlations (for $N_f=0$) unchanged due to translational
invariance together with the universality of the measure.}
 since the determinant can still be
diagonalized by a unitary transformation. The Dirac operator
eigenvalues simply obtained a {\it constant} imaginary contribution, 
$x_j\to x_j+i\muc$, where the $x_j$ are the original real
eigenvalues. This can also be seen from \cite{FK} where a 
more general model with $H+i\Gamma=J$ was studied for $N_f=0$, with
$\Gamma$ being a diagonal matrix. 
The imaginary part of $J$ was fixed with a delta-function constraint in 
the measure. Specifying $\Gamma=\muc \mathbf{1}$ 
immediately leads to a
constant shift of the real eigenvalues into the complex plane. The
equivalence of the delta-measure and the canonical exponential
measure was shown in \cite{AV} in the microscopic limit.

When we take the large-$N$
limit to determine the RMT correlations there are two
possibilities. If we keep $N\muc$ fixed similar to the microscopic 
rescaling of eigenvalues and masses, $Nx_j$ and $Nm_f$ respectively,
we reobtain the known massive correlations \cite{DNII} at shifted
eigenvalues. The second possibility is to keep $\muc$ fixed at
$N\to\infty$ which immediately leads to a complete quenching and thus
to the $N_f=0$ correlations \cite{VZ}.
Although in general a shift of $H$ by a constant matrix known as
random plus deterministic correlations may drastically change the
spectrum (e.g. \cite{FK})
this is not the case here introducing $\muc$ in the most naive way.

In order to incorporate the influence of a chemical potential that
leads to complex eigenvalues we will thus 
pursue another way. Namely we will replace the Hermitian matrix $H$ by 
a complex matrix $J$, without changing further the structure of the Dirac 
operator determinant. Both Hermitian and anti-Hermitian part of the
matrix $J$ will be averaged with a Gaussian measure. Given this
general structure we are left with a few questions. Following
\cite{FKS97} there exist two different large-$N$ limits, the limit of
weak and strong non-Hermiticity. Both limits provide different
correlation functions already in the flavorless case as discussed in
\cite{FKS97}. In the following we will generalize these results in
both limits to
an arbitrary number of massive quark flavors in sections \ref{I} and
\ref{II}. It is clearly the first
limit of weak non-Hermiticity which 
makes contact to the problem posed. In this limit 
there exists a parameter $\al$ which interpolates
between non-Hermitian $(\al>0)$ and Hermitian matrices $(\al=0)$. The
role of $\al$ as a function of $\muc$ will be discussed at least
qualitatively. 
The  microscopic large-$N$ limit we will take slightly differs from 
\cite{FKS97} in the weak non-Hermitian case. Because of the analogue \cite{VZ} 
of the Banks-Casher relation relating the condensate to the
Dirac operator spectral density (per unit length) at the origin 
we are interested in 
the eigenvalues very close to zero. We will therefore define a 
microscopic origin scaling limit in which both  
the real and imaginary part of the complex
eigenvalues $z_j$ of $J$ will be equally rescaled by the mean level
spacing. In the limit $\al\to 0 $ we will then recover the 
correlation functions of the Hermitian model \cite{VZ,DNII} with $2N_f$ 
massless or massive flavors.

The article is organized as follows. In section \ref{general} we
present the general method how to calculate complex eigenvalue 
spectra with massive quarks. 
Here, we follow closely \cite{FKS97} using orthogonal
polynomials in the complex plane and show how to extend their results
to $N_f\neq 0$. In the sequel we then discuss 
the two microscopic large-$N$ limits of weak and strong
non-Hermiticity in sections \ref{I} and \ref{II},
respectively. Section \ref{I} contains a discussion of the
universality of our results and a comparison to the Hermitian case.
We conclude in section \ref{conclude}.

\setcounter{equation}{0}
\section{General result: massive correlation functions at finite-$N$}
\label{general}

We start this section by stating the RMT partition function for
QCD$_3$ with a complex Dirac operator and $2N_f$ massive flavors
\beq
{\cal Z}_{QCD3}^{(2N_f)}(\{m_f\}) \ \sim\ \int dJdJ^{\dagger}
\prod_{f=1}^{N_f}\left|\det[J-im_f]\right|^2 
 \exp\left[-\frac{N}{1-\tau^2}\tr\left(JJ^\dagger-\tau\re
     J^2\right)\right] \ .
\label{Z}
\eeq
Here, $J$ is a complex matrix and we integrate over independent matrix
elements. We have chosen to take the absolute value of the Dirac
operator determinant in order to make the partition function real
valued. It can be written in terms of
complex eigenvalues 
\beq
{\cal Z}_N^{(2N_f)}(\{m_f\}) \ =\ \int \prod_{j=1}^N
\left(d^2\!z_j \prod_{f=1}^{N_f} |z_j-im_f|^2\ w^2(z_j)\right)
 |\Delta_N(z_1,\ldots,z_N)|^2 \ ,
\label{Zev}
\eeq
with the Vandermonde reading $\Delta_N(z_1,\ldots,z_N)=\prod_{k>l}^N
(z_k-z_l)$. The weight function is given by 
\beq
w^2(z)\ =\ \exp\left[-\frac{N}{1-\tau^2}\left(|z|^2 -\frac{\tau}{2} 
(z^2+\zc^2)\right)\right]\ . 
\label{weight}
\eeq 
The diagonalization of $J$ has been described in detail in
\cite{FKS98} for $N_f=0$ and we only repeat here the ingredients
important for the Hermiticity properties.
To justify the form of the average in the partition function eq. (\ref{Z})
we start from the following decomposition of an arbitrary complex
matrix 
\beq
J\ =\ H + i \sqrt{\frac{1-\tau}{1+\tau}} A \ .
\label{Jdef}
\eeq
Here, $H$ and $A$ are both Hermitian matrices with Gaussian weight and
equal variance $(1+\tau)/(2N)$. The parameter $\tau\in[0,1]$ governs
the degree of 
non-Hermiticity. While for $\tau=0$ we have maximal non-Hermiticity for 
$\tau=1$ we obtain back a Hermitian matrix 
$J=H$. Furthermore, $\tau$ controls the
correlations between matrix elements of $J$ as
$<J_{kl}J_{lk}>\ =\tau/N$ and $<J_{kl}J^\ast_{kl}>\ =1/N$.
At large-$N$ we will either keep $1-\tau$ or $N(1-\tau)$ fixed, the
strong or weak non-Hermitian limit.
The product of the Gaussian measures ${\cal P}(H)\sim
\exp[-N/(1+\tau)\tr H^2]$ for the matrices $H$ and $A$  can be simply
rewritten as a single measure for the complex matrix $J$,
${\cal P}(J)\sim \exp[-N/(1-\tau^2)\tr(JJ^\dagger-\tau\re J^2)]$. 
This leads to the appearance of particular 
the weight-function eq. (\ref{weight})
for the complex eigenvalues $z_j$ of $J$.

The crucial observation is that the correlation functions for the
partitions function eq. (\ref{Zev}) can be obtained using the powerful 
technique of orthogonal polynomials. Let us therefore define a set of
polynomials $P_k^{(2N_f)}(z)$ orthonormal in the complex plane $z=x+iy$
by 
\beq
\int d^2\!z \prod_{f=1}^{N_f} |z-im_f|^2\ w^2(z)\ 
P_k^{(2N_f)}(z)P_l^{(2N_f)}(\zc) \ =\ \delta_{kl} \ ,
\label{OP}
\eeq
where $d^2\!z = dxdy$ and $x,y \ \in {\mathbf R}$.
In the case $N_f=0$ these are given by appropriately
rescaled standard Hermite polynomials, $P_k^{(0)}(z)\sim
H_k(z\sqrt{N/\tau})$, an observation made in \cite{FGIL}. Furthermore
we define the $k$-point correlation function for finite-$N$
in the usual way:
\be
R_N^{(2N_f)}(z_1,\ldots,z_k) &=& \frac{N!}{(N-k)!}
\int d^2\!z_{k+1}\ldots d^2\!z_N \ {\cal P}_N(\{z_j\})\ , \nn\\ 
{\cal P}_N(\{z_j\}) &=& \frac{1}{{\cal Z}_N^{(2N_f)}(\{m_f\})}
\prod_{l=1}^N\left(
\prod_{f=1}^{N_f} |z_l-im_f|^2\ w^2(z_l)\right) 
|\Delta_N(z_1,\ldots,z_N)|^2\ . 
\label{Rkdef}
\ee
Using the standard technique of orthogonal polynomials \cite{Mehta} 
the correlation functions can be expressed in terms of the kernel of the 
orthogonal polynomials 
\beq
K^{(2N_f)}_N(z_1,z_2^\ast) \ =\ 
\prod_{f=1}^{N_f} |z_1-im_f||z_2^\ast+im_f|\ 
w(z_1)w(z_2^\ast)\sum_{l=0}^{N-1} 
P_l^{(2N_f)}(z_1)P_l^{(2N_f)}(z_2^\ast)
\label{kernel}
\eeq
as
\beq
R_N^{(2N_f)}(z_1,\ldots,z_k) \ =\
\det_{i,j=1,\ldots,k}\left[K_N^{(2N_f)}(z_i,z_j^\ast)\right]\ .
\label{MM}
\eeq
In order to calculate the correlation functions for an arbitrary
number of massive flavors $2N_f$ we would have to determine the
orthogonal polynomials in eq. (\ref{OP}) and then evaluate the
kernel. This could in principle be done in analogy
to \cite{DNII} starting from Hermite polynomials for
$N_f=0$ and iteratively adding massive flavors (see also Theorem 2.5 in
\cite{Sze}). 
We will proceed in another way by directly relating
the massive correlation functions $R_N^{(2N_f)}$ to the flavorless
ones, $R_N^{(0)}$, as it was done in \cite{AK} for the chiral ensembles 
of real eigenvalues. The advantage of this derivation is twofold. First of
all we will immediately obtain a closed and simple form for all $N_f$,
without calculating intermediate polynomials.
Second, the RMT universality of our
results will be directly inherited given the universality of the
flavorless case can be shown.

We now describe how to obtain the massive correlation
functions. The key observation borrowed from \cite{AK} is to 
incorporate the Dirac
operator determinant with $2N_f$ flavors into a larger Vandermonde
\beq
\prod_{j=1}^N\prod_{f=1}^{N_f} |z_j-im_f|^2\ |\Delta_N(z_1,\ldots,z_N)|^2
\ =\ \frac{|\Delta_{N+N_f}(z_1,\ldots,z_N,im_1,\ldots,im_{N_f})|^2}
{|\Delta_{N_f}(im_1,\ldots,im_{N_f})|^{2}} \ .
\label{delta}
\eeq
Note that due to the pairing of imaginary masses in QCD$_3$ we do not need to
impose an extra degeneracy to the power of the Dyson index $\beta$ 
as it was done in \cite{AK} for the chiral ensembles. Furthermore, the relation
(\ref{delta}) is the technical reason why we have to take the absolute 
value of the complex Dirac operator determinant in eq. (\ref{Z}) to
begin with. Otherwise we would not be able to relate the flavorless
correlations to the massive ones. 
Inserting the identity (\ref{delta}) into the definition (\ref{Rkdef}) we
obtain\footnote{Strictly speaking eqs. (\ref{Rkint})-(\ref{master})
  only hold for finite-$N$ in the normalization where the
  weight function $w(z)$ eq. (\ref{weight}) is 
  $N$-independent (see \cite{AK}). In the large-$N$ limit to 
  be taken later this difference becomes immaterial.\label{N}}
\beq
R_N^{(2N_f)}(z_1,\ldots,z_k) \ =\ \frac{N!}{(N+N_f)!}
\frac{{\cal Z}_{N+N_f}^{(0)}}{{\cal Z}_N^{(2N_f)}(\{m_f\})}
\prod_{f=1}^{N_f}\frac{1}{w^2(im_f)} 
\ \frac{R_{N+N_f}^{(0)}(z_1,\ldots,z_k,im_1,\ldots,im_{N_f})}
{|\Delta_{N_f}(im_1,\ldots,im_{N_f})|^{2}} \ .
\label{Rkint}
\eeq
In order to evaluate the mass-dependent normalization factor 
${\cal Z}_N^{(2N_f)}(\{m_f\})$ we also insert eq. (\ref{delta}) into the
definition (\ref{Zev}) to obtain
\beq
{\cal Z}_N^{(2N_f)}(\{m_f\}) \ =\  \frac{N!}{(N+N_f)!}
\prod_{f=1}^{N_f}\frac{1}{w^2(im_f)} 
\ {\cal Z}_{N+N_f}^{(0)}\frac{R_{N_f}^{(0)}(im_1,\ldots,im_{N_f})}
{|\Delta_{N_f}(im_1,\ldots,im_{N_f})|^{2}} \ . 
\label{Zint}
\eeq
With these two results we obtain the following master formula for the
massive $k$-point correlation function of complex eigenvalues in terms 
of the known flavorless correlation functions of \cite{FKS97}:
\beq
R_N^{(2N_f)}(z_1,\ldots,z_k) \ =\
\frac{R_{N+N_f}^{(0)}(z_1,\ldots,z_k,im_1,\ldots,im_{N_f})}
{R_{N_f}^{(0)}(im_1,\ldots,im_{N_f})} \ .
\label{master}
\eeq
The right hand side is entirely given in terms of the zero-flavor
kernel eq. (\ref{kernel}) from \cite{FKS97} 
which we explicitly display here in terms of
the Hermite polynomials,
\beq
K^{(0)}_N(z_1,z_2^\ast) \ =\ 
w(z_1)w(z_2^\ast)\frac{N}{\pi\sqrt{1-\tau^2}}\sum_{l=0}^{N-1} 
\frac{\tau^n}{n!}
H_l\left(\sqrt{\frac{N}{\tau}}z_1\right)
H_l\left(\sqrt{\frac{N}{\tau}}z_2^\ast\right) \ .
\label{kernelN}
\eeq
Since all arguments on the right hand side of eq. (\ref{master}) are
complex the insertion of the imaginary quark masses $im_f$ does not
constitute an analytical continuation, in contrast to the case of real 
eigenvalues considered in \cite{AK}. We thus do not have to face the
subtleties that occured there when taking the absolute value in the
large-$N$ limit. 
 
Let us add that we can use exactly the same strategy 
when restricting the partition
function eq. (\ref{Zev}) to real eigenvalues. We thus recover without 
any effort from eq. (\ref{master}) 
the results of \cite{DNII} for massive microscopic correlation functions 
out of the flavorless ones. 
In fact we will reobtain their results from our correlation functions
of complex eigenvalues in the Hermitian
limit in subsection \ref{Herm}.
 
While eq. (\ref{master})
establishes an exact result for finite$^{\ref{N}}$-$N$ we will be primarily 
interested in taking the microscopic large-$N$ limit. As it has been
shown \cite{FKS97} there exist two distinct microscopic large-$N$
limits: the limit of weak and strong non-Hermiticity. 
The explicit form of the correlations eq. (\ref{master}) 
in these two limits will be
the subject of the next two sections.

\setcounter{equation}{0}
\section{Microscopic limit I: weak non-Hermiticity}
\label{I}

Let us start by by defining the limit of weak non-Hermiticity
recalling the results of \cite{FKS97,FKS98}. 
We will take the limit $\tau\to 1$ 
in a controlled way in which the matrix $J$ in eq. (\ref{Jdef})
becomes almost Hermitian. Namely we will keep 
\beq
\lim_{N\to\infty}2N(1-\tau)\ \equiv \ \al^2
\label{tau}
\eeq
fixed 
in the limit where $N\to\infty$. The Hermitian limit can then be recovered 
by taking $\al\to0$. From the requirement of keeping the
weight in eq. (\ref{weight}) finite we immediately see that we have to 
rescale the arguments $z_{1,2}$ of the kernel $K_N^{(2N_f)}(z_1,z_2^\ast)$ in
eq. (\ref{kernel}) appropriately. The authors of \cite{FKS97} find
non-trivial correlations when 
\beq
\re(z_1- z_2)={\cal O}(1/N) \ \ \mbox{and}\ \ 
\im\ z_{1,2}={\cal O}(1/N)\ ,
\label{FKSlimit}
\eeq
with $1/N$ being proportional to the mean level spacing while 
$\re\ z_{1,2}={\cal O}(1)$ is kept finite. In other words they require 
$\re\ z \gg \im\ z$ in addition to the limit eq. (\ref{tau}).
Let us furthermore define the mean spectral density of 
the real part of the eigenvalues only:
\beq
\nu(x) \ \equiv\ \frac{1}{N}< \rho(x=\re\ z)>\ \ .
\label{nu}
\eeq
Here, the large-$N$ average is taken without rescaling the eigenvalues 
or unfolding, leading to the 
macroscopic density.
For a Gaussian weight this is just the semi-circle
$\nu(x)=\frac{1}{2\pi}\sqrt{4-x^2}$, since the average over the matrix 
$A$ drops out. In \cite{FKS98} 
the final result for the asymptotic kernel only depends
on the average real part of
the eigenvalues, $X=\re(z_1+z_2)/2$,
through the mean eigenvalue density $\nu(X)$, 
apart from a pre-factor. It reads
\be
K_N^{(0)}(z_1,z_2^\ast) &\stackrel{N\to\infty}{\longrightarrow}& 
\frac{N}{\pi\al}
\exp\left[\frac{i}{2}XN\im(z_1-z_2)\right]
\exp\left[-\frac{N^2}{\al^2}\left(\im^2 z_1+\im^2 z_2^\ast\right)\right] \nn\\
&&\times\int_{-\pi\nu(X)}^{\pi\nu(X)}
\frac{du}{\sqrt{2\pi}}
\exp\left[-\frac{\al^2u^2}{2}+iN(z_1-z_2^\ast)u\right] \ .
\label{FKSkernel}
\ee
After unfolding the kernel, the correlation functions
and the eigenvalues according to eq. (\ref{FKSlimit}) the
zero- and the $N_f$-flavor correlators can be immediately read off from
eqs. (\ref{MM}) and (\ref{master}), respectively.  

Because of the Banks-Casher relation between the condensate and the
mean spectral density at the origin from QCD$_3$ \cite{VZ} 
we are interested in the
microscopic origin scaling limit. We can therefore relax the condition 
$\re\ z\gg\im\ z$ since we rescale both real and imaginary part in
the same way. The microscopic origin scaling limit is defined by
introducing variables
\beq
N(\re\ z+i\im\ z)\ =\ Nz \ \equiv \xi \ \ \ ,\ Nm_f \ \equiv\ \mu_f
\label{micro}
\eeq
which is kept fixed at $N\to\infty$. In addition we have rescaled the masses
accordingly since they appear on the same footing in the determinant
of the partition function eq. (\ref{Zev}). 
In consequence the kernel
eq. (\ref{FKSkernel}) simplifies furthermore as now the average $X$ is also
quantity of order ${\cal O}(1/N)$ vanishing at large-$N$. We thus 
obtain for the zero-flavor kernel of weakly non-Hermitian eigenvalues
\beq
K_S^{(0)}(\xi_1,\xi_2^\ast) \ =\ 
\frac{1}{\pi\al}\exp\left[-\frac{1}{\al^2}
\left(\im^2\xi_1+\im^2\xi_2^\ast\right)\right]
g_\al(\xi_1-\xi_2^\ast) \ ,
\label{newkernel}
\eeq
where\footnote{Note the difference by a factor of $\frac{i}{2}$ in the
  definition compared to \cite{FKS97}.}
\beq
g_{\al}(\xi) \ \equiv\ \int_{-\pi\nu(0)}^{\pi\nu(0)}
\frac{du}{\sqrt{2\pi}}\exp\left[-\frac{\al^2u^2}{2}+i\xi u\right] \ .
\label{gdef}
\eeq
Note that $g_{\al}(\xi)$ is an even function in $\xi$ due to the range 
of integration.
Although the result eq. (\ref{FKSkernel}) was obtained with a Gaussian 
potential only, with $\pi\nu(0)=1$, we keep the dependence on
$\nu(0)$. It directly connects to the condensate at zero chemical
potential and in the Hermitian limit it plays the role
of a universal parameter. The issue of universality will be discussed
separately below. 

The result for the zero-flavor $k$-point correlation functions
simplifies as the exponential 
pre-factors of the kernel can be taken out of the determinant. 
Defining the rescaled or unfolded correlation functions, 
\beq
\rho^{(0)}_S(\xi_1,\ldots,\xi_k)\ \equiv\ 
\lim_{N\to\infty} N^{-k} R_N^{(0)}(N^{-1}\xi_1,\ldots,N^{-1}\xi_k)
\label{rhodef}
\eeq
we obtain from inserting eq. (\ref{newkernel}) into eq. (\ref{MM})
\beq
\rho^{(0)}_S(\xi_1,\ldots,\xi_k)\ =\ \left(\frac{1}{\pi\al}\right)^k 
 \prod_{i=1}^{k}\exp\left[-\frac{2}{\al^2}\im^2\xi_i\right]
\det_{1\leq j,l\leq k}\left[g_\al(\xi_j-\xi_l^\ast)\right] \ , 
\label{rhok}
\eeq
where we have used that $\im^2\xi=\im^2\xi^\ast$.
This also gives us the building blocks for the denominator in
eq. (\ref{master}) with purely imaginary arguments. 
We can now read off the massive correlation function using the
zero-flavor kernel eq. (\ref{newkernel}) in eq. (\ref{master})
\beq
\rho^{(2N_f)}_S(\xi_1,\ldots,\xi_k)\ =\  
\frac{
{
\begin{array}{c}\det\\ 
\mbox{{\footnotesize $1\leq j,l\leq k$}}\\
\mbox{{\footnotesize $1\leq f,h\leq N_f$}}\end{array}}
\left[
\begin{array}{ll}
K_S^{(0)}(\xi_j,\xi_l^\ast) &K_S^{(0)}(\xi_j,-i\mu_h) \\
& \\
K_S^{(0)}(i\mu_f,\xi_l^\ast)&K_S^{(0)}(i\mu_f,-i\mu_h) 
\end{array}\right]}
{\det_{1\leq f,h\leq N_f}[K_S^{(0)}(i\mu_f,-i\mu_h)]} \ .
\label{rhomkernel}
\eeq
The first main result of this article, the massive 
$N_f$-flavor $k$-point correlation function is thus reading
\beq
\rho^{(2N_f)}_S(\xi_1,\ldots,\xi_k)= \left(\frac{1}{\pi\al}\right)^k 
 \prod_{i=1}^{k}\exp\left[-\frac{2}{\al^2}\im^2\xi_i\right]
\frac{
{
\begin{array}{c}\det\\ 
\mbox{{\footnotesize $1\leq j,l\leq k$}}\\
\mbox{{\footnotesize $1\leq f,h\leq N_f$}}\end{array}}
\left[
\begin{array}{ll}
g_\al(\xi_j -\xi_l^\ast)& g_\al(\xi_j +i\mu_h) \\ 
& \\
g_\al(i\mu_f-\xi_l^\ast)& g_\al(i\mu_f+i\mu_h)
\end{array}\right]}
{\det_{1\leq f,h\leq N_f}[g_\al(i\mu_f+i\mu_h)]}.
\label{rhomk}
\eeq
We mention as an aside that from the connected part of the two-point
function we can determine the absolute value squared of the
corresponding {\it massive} kernel, $|K_S^{(2N_f)}(\xi_1,\xi_2^\ast)|^2$. 
This follows from the definition
(\ref{MM}).

\begin{figure}
\centerline{\epsfig{figure=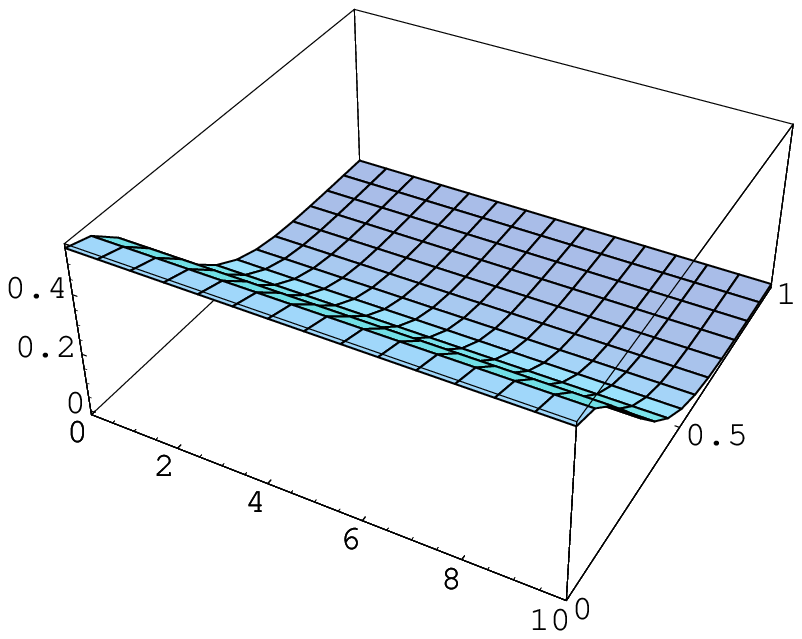,width=20pc
}
\put(-160,20){$\re\ \xi$}
\put(-40,40){$\im\ \xi$}
\put(-240,140){$\rho_s^{(0)}(\xi)$}
\put(230.1,0.1){$\xi$}
\put(5.1,160.1){$\rho_s^{(2)}(\xi)$}
\epsfig{figure=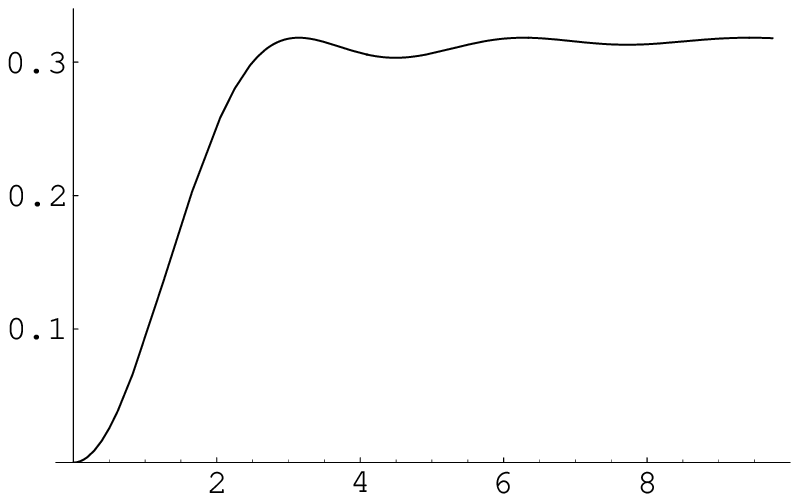,width=20pc}}
\caption{
The flavorless density in the complex plane $\rho_s^{(0)}(\xi)$ 
at $\al^2=0.2$ (left) and the 
microscopic density of real eigenvalues for $2N_f=2$ massless flavors (right).
}
\label{Nf=0}
\end{figure}

In the following we will give a few explicit examples and discuss the
limit of massless flavors. For $N_f=0,1$ the microscopic density reads 
\be
\rho^{(0)}_S(\xi)      &=& \frac{1}{\pi\al}
\exp\left[-\frac{2}{\al^2}\im^2\xi\right]g_\al(2i\im\ \xi) \ ,\label{rho0}\\
\rho^{(2)}_S(\xi) &=& \frac{1}{\pi\al}
\exp\left[-\frac{2}{\al^2}\im^2\xi\right]\left\{ g_\al(2i\im\ \xi)
-\frac{g_\al(\xi+i\mu)g_\al(i\mu-\xi^\ast)}{g_\al(2i\mu)}\right\} \ .
\label{rho1}
\ee
While the flavorless result eq. (\ref{rho0}) 
follows from the origin scaling limit of \cite{FKS96} 
the $N_f=1$ flavor case is new. The 
massless limit is easily obtained by setting $\mu=0$ in
eq. (\ref{rho1}). We have plotted in Figs. \ref{Nf=0} and \ref{Nf=1} 
both cases $N_f=0$ and $1$, respectively,  
for a comparison to the known Hermitian limit.

With the general structure of the massive correlators being clear from
eq. (\ref{rhomk}) we also give explicitly the massless two-flavor case:
\be
\rho^{(4)}_S(\xi) &=& \frac{1}{\pi\al}
\exp\left[-\frac{2}{\al^2}\im^2\xi\right]\left\{ g_\al(2i\im\ \xi)
\ -\ \frac{g_\al(\xi)g_\al(\xi^\ast)}{g_\al(0)} \right.\nn\\
&&\left.+\ \frac{
\left(\sqrt{\frac{2}{\pi}}\mbox{e}^{-\al^2/2}\sin\xi -\xi g_\al(\xi)\right)
\left(\sqrt{\frac{2}{\pi}}\mbox{e}^{-\al^2/2}\sin\xi^\ast
-\xi^\ast g_\al(\xi^\ast)\right)}
{\left(\sqrt{\frac{2}{\pi}}\mbox{e}^{-\al^2/2}-g_\al(0)\right)}\right\} .
\label{rho2}
\ee
In contrast to the correlation functions of real eigenvalues \cite{VZ} 
the massless correlation functions do not simplify here. 
The procedure of sending successively all masses to zero in
eq. (\ref{rhomk}) leading to higher order derivatives of $g_\al$ will
not simplify the determinant structure. The reason for this will become 
clear when comparing to the Hermitian limit below. 
Taking for example the microscopic density with $N_f$ massless flavors 
we will obtain a determinant of size $N_f+1$ even in the limit $\al\to 0$. 
It is only through non-trivial identities that this expression can be
rewritten as a sum of bilinear combinations of half-integer
Bessel-functions for an arbitrary Number $2N_f$ 
of massless flavors as first derived in \cite{VZ}.

\subsection{The Hermitian limit}
\label{Herm}

We will now discuss the Hermitian limit when $\al\to0$. In the
simplest case of zero flavors the microscopic density of the
Hermitian matrix model is simply constant.
For the complex density eq. (\ref{rho0}) 
this feature is maintained on the real axis while the density smoothly 
spreads into the complex plane as depicted in Fig. \ref{Nf=0} (left) (see
also \cite{FKS96}). Taking the limit  $\al\to0$ of eq. (\ref{rho0}) we 
obtain a delta-function times a constant, the density of the real
eigenvalues:
\beq
\lim_{\al\to0}\rho_S^{(0)}(\xi)\ =\ \frac{1}{\sqrt{\pi}} 
\delta\left(\sqrt{2}\im\ \xi\right)
 \lim_{\al\to0}g_\al(2i\im\ \xi) \ =\ \nu(0)  \delta\left(\im\ \xi\right)\ ,
\label{rho0herm}
\eeq
where in the last step we have used
$\lim_{\al\to0}g_\al(0)=\sqrt{2\pi}\nu(0)$. If we unfold the
microscopic density as $\rho_S(\xi)=\lim_{N\to\infty}D R_N(D\xi)$
with respect to the full mean level spacing,
$D=1/(\pi\nu(0)N)$, we obtain the parameter free result
$\rho_S^{(0)}(\xi)=1/\pi$ \cite{VZ}.
Next, we turn to $2N_f=2$ where we take the massless limit for
simplicity. In the Hermitian model the density is given by \cite{VZ}
\beq
\rho_S^{(2)}(\xi)\ =\ \frac1\pi \left(1-\frac{\sin^2(\xi)}{\xi^2}\right)
\ ,\ \ \ \ \xi\ \in {\mathbf R} \ \ ,
\label{rho1herm}
\eeq

\begin{figure}[-h]
\centerline{\epsfig{figure=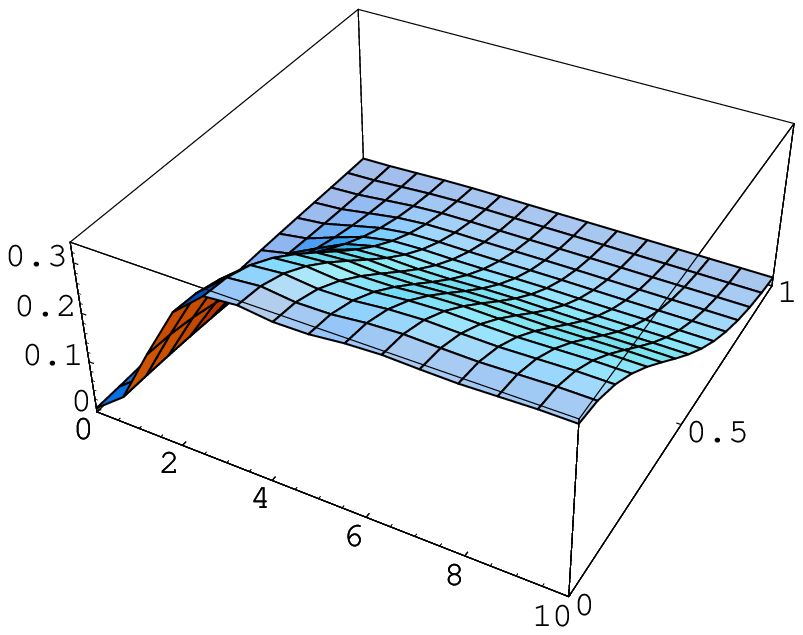,width=20pc}
\put(-160,20){$\re\ \xi$}
\put(-40,40){$\im\ \xi$}
\put(-240,140){$\rho_s^{(2)}(\xi)$}
\put(80,20){$\re\ \xi$}
\put(200,40){$\im\ \xi$}
\put(0,140){$\rho_s^{(2)}(\xi)$}
\epsfig{figure=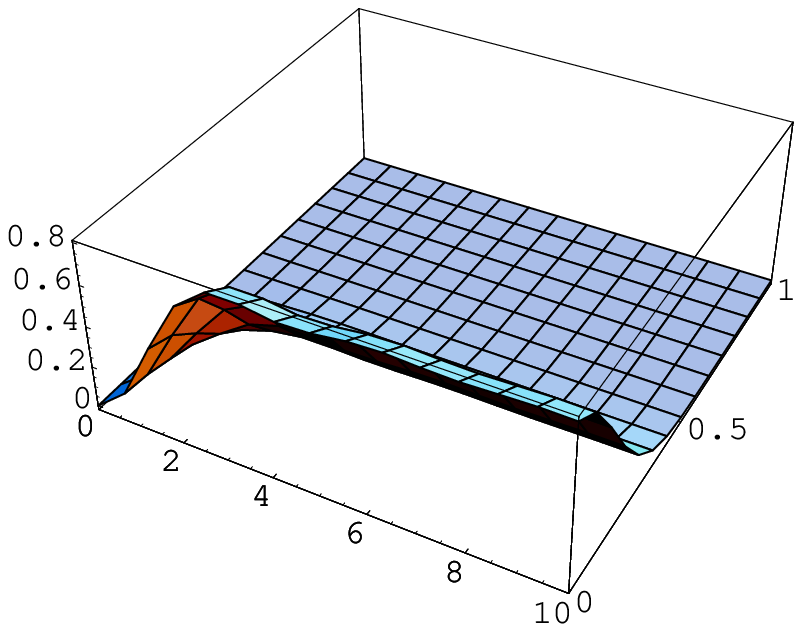,width=20pc}}
\caption{
The microscopic density in the complex plane $\rho_s^{(2)}(\xi)$ 
of $2N_f=2$ massless flavors for different values of $\al^2=0.5$ (left) and
$\al^2=0.1$ (right).
}
\label{Nf=1}
\end{figure}

\noindent
which is plotted in Fig. \ref{Nf=0} (right). The corresponding density in 
the complex plane, 
eq. (\ref{rho1}) with $\mu=0$, is plotted in Fig. \ref{Nf=1} for two
different values of $\al$. It very nicely combines 
both properties of Fig. \ref{Nf=0}, the oscillations of  
eq. (\ref{rho1herm}) on the real axis (right)
and the same spreading into the complex plane (left) of eq. (\ref{rho0}) for
$N_f=0$.  For decreasing $\al$ the distribution
gets narrower, slowly approaching the delta-function. This is also the 
reason for the different normalization in Figs. \ref{Nf=0} and
\ref{Nf=1} for different values of $\al$.
More mathematically we obtain from eq. (\ref{rho1})
\beq
\lim_{\al\to0}\rho_S^{(2)}(\xi)\ =\ 
\nu(0)\delta(\im\ \xi)
\left(1-\frac{\sin^2(\pi\nu(0)\xi)}{(\pi\nu(0)\xi)^2}\right)
\ .
\label{rho1lim}
\eeq
When unfolding with respect to the mean level spacing $D$ we thus
recover eq. (\ref{rho1herm}).

Let us now turn to the Hermitian limit in the 
general case. For that purpose we first
consider the Hermitian limit of the zero-flavor Kernel
eq. (\ref{newkernel}) as a building block. We obtain \cite{FKS97} 
\beq
\lim_{\al\to0}
K_S^{(0)}(\xi_1,\xi_2^\ast) \ =\ 
\nu(0)\sqrt{2}\ 
\delta\left(\sqrt{\im^2\xi_1+\im^2\xi_2}\right)
\frac{\sin(\pi\nu(0)(\xi_1-\xi_2^\ast))}{\pi\nu(0)(\xi_1-\xi_2^\ast)} \ .
\label{sine}
\eeq
After unfolding with respect to $D$ instead of $1/N$ this becomes
precisely the parameter free universal sine-kernel of \cite{BZ} on the 
real axis. It
therefore does not come as a surprise that we also recover the known
result for the microscopic massive $k$-point correlations 
\be
\lim_{\al\to0}
\rho^{(2N_f)}_S(\xi_1,\ldots,\xi_k) &=& \nu(0)^k 
 \prod_{i=1}^{k}\delta\left(\im\ \xi_i\right)     \label{rhomklim}
\\
&&\times\frac{
{\begin{array}{c}\det\\ 
\mbox{{\footnotesize $1\leq j,l\leq k$}}\\
\mbox{{\footnotesize $1\leq f,h\leq N_f$}}
\end{array}
}
\left[
\begin{array}{ll}
\displaystyle
\frac{\sin(\pi\nu(0)(\xi_j - \xi_l))}{\pi\nu(0)(\xi_j -\xi_l)} & 
\displaystyle
\frac{\sin(\pi\nu(0)(\xi_j +i\mu_h))}{\pi\nu(0)(\xi_j +i\mu_h)} \\
& \\
\displaystyle
\frac{\sin(\pi\nu(0)(i\mu_f- \xi_l))}{\pi\nu(0)(i\mu_f-\xi_l)} &
\displaystyle
\frac{\sinh(\pi\nu(0)(\mu_f +\mu_h))}{\pi\nu(0)(\mu_f +\mu_h)}
\end{array}\right]}
{\det_{1\leq f,h\leq N_f}\left[
\displaystyle
\frac{\sinh(\pi\nu(0)(\mu_f + \mu_h))}{\pi\nu(0)(\mu_f +\mu_h)}\right]} \ .
\nn
\ee
Taken in units of $D$ it exactly coincides with the corresponding
result for the Hermitian model as derived in \cite{Szabo} from the
underlying field theory. The equivalence with the original form
\cite{DNII} can be shown using consistency conditions on QCD$_3$ finite volume 
partition functions \cite{ADII}.

Let us come back to the remark at the end of section \ref{general}. 
Had we taken the partition function
eq. (\ref{Zev}) for real eigenvalues in the beginning, together with a
general polynomial in the exponential of the weight function
eq. (\ref{weight}), we would have immediately arrived at
eq. (\ref{rhomklim}) after employing eq. (\ref{master}) and the
zero-flavor kernel eq. (\ref{sine}). We therefore rederive the
results of \cite{DNII} without any effort. The universality of our
results then follows from that of the sine-kernel \cite{BZ,ADMN}.

\subsection{Universality}

Up to now we have only discussed a large-$N$ limit in which the
eigenvalues are rescaled with the mean level spacing, the microscopic
limit. When we want to discuss the issue of universality we have to
distinguish another type of large-$N$ limit, the macroscopic limit. In 
this limit the short range fluctuations typically of the order $1/N$ are
smoothed and the macroscopic (or smoothed or wide) correlation
functions are obtained. Technically speaking the large-$N$ limit is
taken without a rescaling (or unfolding) of the eigenvalues. This
limit is also meaningful since here universal
correlation functions can be obtained \cite{AJM} as well. An example for such
a macroscopic correlation function is the density in eq. (\ref{nu}), the
smoothed eigenvalue density of the real part of the eigenvalues. For
all three Gaussian ensembles of real eigenvalues it is given by the
Wigner semi-circle. However, for a measure including terms of higher order
than Gaussian the macroscopic 
spectral density $\nu(x)$ itself is non-universal as
it explicitly includes all parameters of the higher order terms.  

Going back to complex correlation functions 
the authors of \cite{FKS98} have shown a certain degree of
universality for the macroscopic spectral density $\rho(x,y)$, with
$z=x+iy$, in
the limit of weak non-Hermiticity. Using supersymmetry they have shown that
it only depends on the real part $x$ trough the combination $\nu(x)$,
eq. (\ref{nu}).
Hence the macroscopic density as a function of $y$ only, denoted by
$\rho_x(y)$ in \cite{FKS98}, is universal in the 
sense that when the scale is set by fixing the real part of the eigenvalues
as $x=X$ in \cite{FKS98} or $x=0$ as in 
our case of origin scaling, $\nu(x)$ plays the role of a 
universal parameter in the case of a non-Gaussian weight function. 
Similar results have been obtained in \cite{Ef} for weakly
non-symmetric real matrices.

A more interesting point is the universality of the fluctuations in
the microscopic correlations we have calculated in eq. (\ref{rhomk}). 
We have shown that given the flavorless correlations determined in
\cite{FKS97} are universal their universality would automatically
carry over to our massive results trough eq. (\ref{master}) in the
microscopic limit. From the derivation \cite{FKS98} of the
kernel eq. (\ref{FKSkernel}) it is obvious, that the occurrence of the
mean spectral density of the real eigenvalues $\nu(X)$ inside the
integral eq. (\ref{FKSkernel}) is strictly related to the form of the
measure eq. (\ref{weight}). We can therefore only {\it conjecture}
that the kernels eq. (\ref{FKSkernel}) and (\ref{newkernel}) and thus
the corresponding correlations are universal also for a more general
weight than eq. (\ref{weight}). 
However, we can offer a non-trivial check in the
Hermitian limit $\al\to 0$. As we have seen in this limit the 
flavorless kernel eq. (\ref{FKSkernel}) maps to
the sine-kernel eq. (\ref{sine}) which is guaranteed to be universal
\cite{BZ,ADMN}. In the microscopic origin scaling limit 
the universal parameter $\nu(0)$ occurs in
the correct place after taking the limit $\al\to 0$ of eq. (\ref{newkernel}).
The same argument thus holds for the Hermitian limit
eq. (\ref{rhomklim}) of eq. (\ref{rhomk}).
It would be highly desirable to repeat the universality proof of
\cite{ADMN} for orthogonal polynomials in the complex plane in order to obtain
eq. (\ref{newkernel}) and consequently eq. (\ref{rhomk}) 
for an for an arbitrary weight function. 

For a very recent discussion of universality of complex spectra from
the point of view of the Fokker-Planck equation we refer to \cite{Shu}.

\subsection{Discussion of 
$\al$ as a function of $\mu_{chem}$}

As it has been discussed already in the introduction the naive
introduction of a chemical potential by writing the Dirac operator as
$H+i\muc$ does not lead to nontrivial correlations of complex
eigenvalues. It either
lead to a complete quenching or, after rescaling with keeping $N\muc$
fixed, to a constant shift in the eigenvalues. 
We therefore have introduced a truly complex Dirac operator $J$ and we
would like to understand the role of the parameter $\al$ 
governing the non-Hermiticity as a function of chemical potential $\muc$. It
would be therefore very instructive to compare to the RMT model in
four dimensions \cite{Steph} and identify the
parameters. However, in QCD$_4$ much less is known. Only the
macroscopic density and its support are known \cite{Steph}
as a functions of $\muc$ 
in the RMT model while the microscopic correlations to be compared with
lattice simulations are still lacking. 

What we can do is to calculate the variation of the imaginary part of
the eigenvalues, $\im\ \xi$, in the limit $\al\to 0$ and compare it to a
vanishing potential $\muc\to 0$ in \cite{Steph}. 
For simplicity let us consider the zero-flavor case with microscopic
density eq. (\ref{rho0}). 
We can calculate the second moment along the imaginary axis 
$ <\im^2\xi>\equiv\int_{-\infty}^\infty\! d\im \xi \
\im^2\xi\ \rho_S^{(0)}(\xi)$ given by 
\beq
<\im^2\xi> \ =\ \frac{1}{4\pi} \al^2 \left(1+\frac23 \al^2\right) .
\label{moment}
\eeq
On the QCD$_4$ side we can ask with which power of $\muc$ the {\it
  unscaled} imaginary part\footnote{Note that in \cite{Steph} for
  vanishing $\muc$ the Dirac eigenvalues are chosen to be purely
  imaginary.} of the eigenvalues vanishes on the boundary of the
support. It follows from \cite{Steph} that
\beq
\im\ z \ =\ \muc^2 \sqrt{4-\re^2z} \ +\ {\cal O}(\muc^4) \ .
\eeq
We notice that the mean spectral density $\nu(x)$ naturally occurs.
It is therefore tempting to identify at least to lowest order
\beq
 \al (\muc) \ \sim\ \muc^2 \ . 
\label{almuc}
\eeq
Of course we do not have to make this identification. We could also
simply take $\al$ as a fit parameter to compare with numerical data
for the QCD$_3$ Dirac operator with chemical potential once they are
available. The analytical 
form of the microscopic correlations would still be fixed by
eq. (\ref{rhomk}). 

Let us mention a direct application for QCD$_4$ with chemical
potential. It has been found empirically that also away from the
origin the bulk correlations of Dirac operator eigenvalues can be very 
well described by RMT \cite{HV} (for the most precise tests we refer to
\cite{GMMW}). In the bulk of the spectrum the Dirac determinant and thus 
the chiral properties are no longer seen
which makes our non-chiral RMT eq. (\ref{Zev}) 
applicable. Thus we can
take the zero-flavor microscopic correlation function of complex
eigenvalues eq. (\ref{rhok}) to test QCD$_4$ correlation functions
with chemical potential in the bulk of the spectrum. As it was shown
in \cite{Steph} the macroscopic spectral density in the complex plane 
is independent of the real part $\re\ z$ of the eigenvalues. 
This is consistent with the quenched microscopic density
eq. (\ref{rho0}). One could thus 
check if the complex eigenvalues decay microscopically as 
depicted in Fig. \ref{Nf=0} (left) along a fixed line of of constant
$\re\ z$ parallel to the real axis.  
Comparing for different values of $\muc$
would then fix the constant $\al$ as a function of $\muc$. We
stress that also the higher order correlation functions are available
analytically in eq. (\ref{rhok}). The corresponding real correlations 
eq. (\ref{rhomklim}) have been already tested in
\cite{GMMW} for vanishing chemical potential.

\setcounter{equation}{0}
\section{Microscopic limit II: strong non-Hermiticity}
\label{II}

In this section we treat the large-$N$ limit in which the parameter
$\tau$, or more specifically the combination
$\sqrt{(1-\tau)/(1+\tau)}$ in the definition (\ref{Jdef}) is kept
fixed. Here, there will be no limit possible to recover the correlations of the
original Hermitian model and thus QCD$_3$ with real eigenvalues, such 
as discussed in subsection \ref{Herm}. However, we will find nontrivial
generalizations of the original work of Ginibre \cite{Gin}. It turns
out, that when introducing quark flavors the microscopic 
spectral density starts
to decay compared to the zero-flavor case, where it is entirely flat
on the support \cite{Gin} (see eq. (\ref{rho0strong})). 

The major difference to the weak non-Hermitian limit comes from the
fact that the mean level spacing is changed to be $D\sim 1/\sqrt{N}$. 
When looking at the weight function eq. (\ref{weight}) with $1-\tau$
fixed we can define the microscopic origin 
scaling limit for strong non-Hermiticity as 
\beq
\sqrt{N}(\re\ z+i\im\ z)\ =\ \sqrt{N}z \ \equiv \ \xi \ \ \ ,\ 
\sqrt{N}m_f \ \equiv\ \mu_f\ .
\label{stronglim}
\eeq
The microscopic kernel which is now defined as 
$K_S(\xi_1,\xi_2^\ast)=K_N(\xi_1/\sqrt{N},\xi_2^\ast/\sqrt{N})/\sqrt{N}$ can 
be directly obtained for $N_f=0$
by taking the limit (\ref{stronglim}) of the result for finite-$N$, 
eq. (\ref{kernelN}) \cite{FKS98}:
\beq
K_S^{(0)}(\xi_1,\xi_2^\ast) \ =\ \frac{1}{\pi(1-\tau^2)}
\exp\left[\frac{1}{2(1-\tau^2)}
\left( 2\xi_1\xi^\ast_2 -|\xi_1|^2-|\xi_2|^2
+\frac{\tau}{2}({\xi^\ast_1}^2 -\xi_1^2 +\xi_2^2
-{\xi^\ast_2}^2)\right)
\right] \ .
\label{kernelstrong}
\eeq
At equal arguments we thus obtain a constant microscopic density for
zero flavors \cite{Gin,FKS98}
\beq
\rho^{(0)}_S(\xi)      \ =\ \frac{1}{\pi(1-\tau^2)} \ .
\label{rho0strong}
\eeq
It coincides with the smoothed or macroscopic spectral density of the 
unscaled eigenvalues \cite{Gin} which has been reobtained in
\cite{FKS98} using the supersymmetric method
\beq
\rho(z)\ =\ \lim_{N\to\infty}\frac1N R_N(z) \ =\ \left\{ 
\begin{array}{cl} 
\displaystyle
\frac{1}{\pi(1-\tau^2)}\ , & 
\displaystyle
\ \ \mbox{if} \ \ \frac{\re^2 z}{(1+\tau)^2}+
\frac{\im^2 z}{(1-\tau)^2}\leq 1 \\
\displaystyle 
0\ , & 
\displaystyle
\ \ \mbox{otherwise}
\end{array}\ .
\right. 
\label{rhomacro}
\eeq
The latter is independent of the number of quarks present as it is obtained
from saddle point analysis where the quark determinants are
subleading. 
We recall that in the RMT of QCD$_4$ 
\cite{Steph} the  support of the macroscopic density 
is as well given by an algebraic curve and, although not being flat,
the density is independent of the real part
of the Dirac operator eigenvalues.

Next we turn to the microscopic correlation functions in the
presence of $2N_f$ massive quarks. Inserting the zero-flavor kernel
eq. (\ref{kernelstrong}) 
into eq.  (\ref{rhomkernel}) now in
the strong non-Hermitian limit eq. (\ref{stronglim}) 
we obtain for the general $k$-point function
\beq
\rho^{(2N_f)}_S(\xi_1,\ldots,\xi_k) = 
\left(\frac{1}{\pi(1-\tau^2)}\right)^k 
 \prod_{i=1}^{k}\exp\left[\frac{-|\xi_i|^2}{1-\tau^2}\right] 
\frac{
{\begin{array}{c}\det\\ 
\mbox{{\footnotesize $1\leq j,l\leq k$}}\\
\mbox{{\footnotesize $1\leq f,h\leq N_f$}}
\end{array}
}
\left[
\begin{array}{ll}
\displaystyle
\exp\left[\frac{\xi_j\ \xi_l^\ast}{1-\tau^2}\right] &
\displaystyle
\exp\left[\frac{-\xi_j\ i\mu_h}{1-\tau^2}\right]     \\
& \\
\displaystyle
\exp\left[\frac{i\mu_f\ \xi_l^\ast}{1-\tau^2}\right] &   
\displaystyle
\exp\left[\frac{\mu_f\ \mu_h}{1-\tau^2}\right] \\
\end{array}\right]}
{\det_{1\leq f,h\leq N_f}\left[
\displaystyle
\exp\left[\frac{\mu_f\ \mu_h}{1-\tau^2}\right]\right]}.
\label{rhomklimstrong}
\eeq
Here, we have taken those parts in eq. (\ref{kernelstrong}) which 
factorize out of the determinant, leading to a partial cancelation.
Let us explicitly display the simplest examples, the density with one
massive flavor  
\beq
\rho^{(2)}_S(\xi) \ =\  \frac{1}{\pi(1-\tau^2)}
\left\{1\ -\ \exp\left[-\frac{1}{1-\tau^2}|\xi-i\mu|^2\right]\right\} \ ,
\label{rho1strong}
\eeq
and the density with two massless flavors
\beq
\rho^{(4)}_S(\xi) \ =\ \frac{1}{\pi(1-\tau^2)}
\left\{1\ -\ \left(1+\frac{|\xi|^2}{1-\tau^2}\right)
\exp\left[-\frac{|\xi|^2}{1-\tau^2}\right]\right\} \ .
\label{rho2strong}
\eeq

\begin{figure}[-h]
\centerline{\epsfig{figure=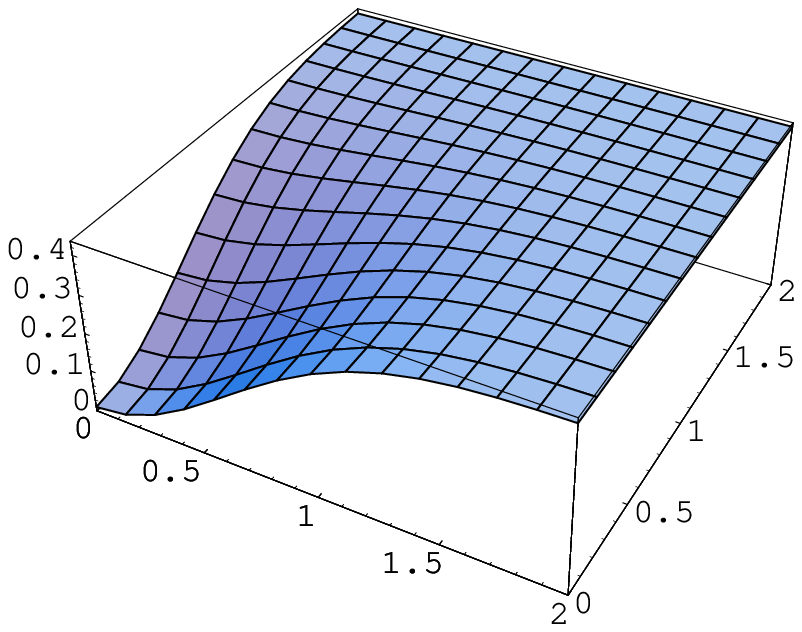,width=20pc}
\put(-160,20){$\re\ \xi$}
\put(-30,50){$\im\ \xi$}
\put(-240,140){$\rho_s^{(2)}(\xi)$}
\put(80,20){$\re\ \xi$}
\put(210,50){$\im\ \xi$}
\put(0,140){$\rho_s^{(4)}(\xi)$}
\epsfig{figure=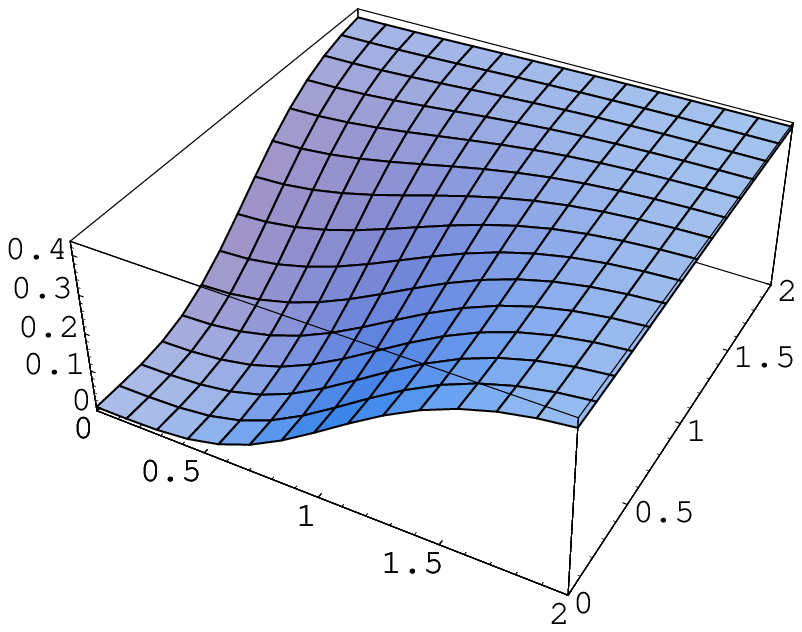,width=20pc}}
\caption{
The massless microscopic density in the complex plane $\rho_s^{(2N_f)}(\xi)$ 
for $2N_f=2$ (left) and $2N_f=4$ (right), both at $\tau=1/2$.
}
\label{Nf=12strong}
\end{figure}

\noindent
In both cases it is no longer constant. 
For vanishing masses both vanish  at the origin $\xi=0$ due to the level
repulsion of the additional quark flavors. For large arguments they
both approach the value of the macroscopic density
eq. (\ref{rhomacro}). As we can see in Fig. \ref{Nf=12strong} for two
flavors (right) the level repulsion is stronger.

We finally note that the microscopic correlations in the 
limit of strong non-Hermiticity can be obtained from those in the weak 
limit by taking $\al\to\infty$, as being mentioned already in
\cite{FKS98}. In this limit eq. (\ref{rhomklimstrong}) follows from
eq. (\ref{rhomk}) when identifying $\al^2\equiv(1-\tau^2)$.
In the simplest example eq. (\ref{rho0}) leads to the constant density 
eq. (\ref{rho0strong}).
We thus find a crossover from the regime of weak to strong non-Hermiticity with
growing $\al$. A similar crossover has been seen for the nearest
neighbor distribution in quenched QCD$_4$ lattice data \cite{MPW} as a
function of growing $\muc$.

\section{Conclusions}
\label{conclude}

We have derived analytic expressions for all correlation functions of
complex Dirac operator eigenvalues in the presence of an arbitrary
even number of massive quark flavors. Here, we have replaced the Dirac
operator for QCD$_3$ by a general complex matrix as we have argued
that a shift of an originally Hermitian operator by a constant
chemical potential, $i\muc$, would not lead 
to non-trivial correlations. We have 
solved the problem in the limit of weak and strong non-Hermiticity and 
have identified the parameter for non-Hermiticity $\al$ as modeling
the influence of a chemical potential. 
Our results may also be useful 
for QCD lattice data with chemical potential 
in four dimensions in the bulk of the spectrum.
In the limit of a Hermitian
Dirac operator we could nicely reproduce the known results for 
real eigenvalues. Moreover, we could offer a very elegant
alternative of calculating massive real eigenvalue correlations and
proving their RMT universality.

There are several open questions we have not addressed so far. The
main task is a generalization of our results to chiral RMT as an
effective model for QCD$_4$ spectra close to the origin. The crucial
step would be to find the corresponding orthogonal polynomials in the
complex plane to make our techniques applicable. 
An open problem within our calculations is the question of
universality of the correlation functions which we could only
conjecture. The problem is to show that 
for an arbitrary weight function we will obtain orthogonal polynomials 
with the same asymptotic as the Hermite polynomials. Given the success
for real eigenvalues in this respect progress should be
possible. Another question is to find the finite-volume partition
function corresponding to the partition function of complex
eigenvalues. In other words one would have to find an appropriate
unitary group integral given by the determinant of the massive 
kernel of complex eigenvalues.

Another open problem would be the generalization to the other
invariant random matrix ensembles, the orthogonal and symplectic ensemble, to
incorporate complex eigenvalues. First results have been obtained only 
for the macroscopic spectral density of almost symmetric matrices using the
supersymmetric method.

\noi\noindent
{\bf Acknowledgments}\\
I wish to thank S. Pepin for useful discussions and reading the manuscript.
This work was supported in part by EU TMR grant no.
ERBFMRXCT97-0122.


\begin{thebibliography}{XX}

\bibitem{HN}N. Hatano and D.R. Nelson,  Phys. Rev. Lett. {\bf 77} (1996)
  570-573 [cond-mat/9603165].

\bibitem{Haa}F. Haake, {\it Quantum Signatures of Chaos}, Springer
  Verlag, Berlin 1991.

\bibitem{GHS}R. Grobe, F. Haake and H.-J. Sommers,
  Phys. Rev. Lett. {\bf 61} (1988) 1899-1902. 

\bibitem{Ef}K.B. Efetov, Phys. Rev. Lett. {\bf 79} (1997) 491-494 
[cond-mat/9702091].

\bibitem{FK}Y.V. Fyodorov and B.A. Khoruzhenko,
Phys. Rev. Lett. {\bf 83} (1999) 65-68 [cond-mat/9903043].

\bibitem{Steph} M.A. Stephanov, Phys. Rev. Lett. {\bf 76} (1996)
  4472-4475 [hep-th/9604003].

\bibitem{latt}S. Chandrasekharan,  Nucl. Phys. Proc. Suppl. {\bf 94}
  (2001) 71-78 [hep-lat/0011022];\newline
F. Karsch, Nucl. Phys. Proc. Suppl. {\bf 83} (2000) 14-23 [hep-lat/9909006].

\bibitem{VW} J.J.M. Verbaarschot and T. Wettig, 
Ann. Rev. Nucl. Part. Sci. {\bf 50} (2000) 343-410 [hep-ph/0003017].

\bibitem{FKS97}Y.V. Fyodorov, B.A. Khoruzhenko and H.-J. Sommers,
  Phys. Rev. Lett. {\bf 79} (1997) 557-561 [cond-mat/9703152].

\bibitem{Gin}J. Ginibre, J. Math. Phys. {\bf 6} (1965) 440-449.

\bibitem{MPW}H. Markum, R. Pullirsch and T. Wettig, 
 Phys. Rev. Lett. {\bf 83} (1999) 484-487 [hep-lat/9906020].

\bibitem{Janik} R.A. Janik, M.A. Nowak, G. Papp and I. Zahed, 
Phys. Rev. Lett. {\bf 81} (1998) 264-267 [hep-ph/9803289].

\bibitem{VZ}J.J.M. Verbaarschot and I. Zahed, Phys. Rev. Lett. {\bf 73} 
(1994) 2288-2291 [hep-th/9405005].

\bibitem{HV} M.A. Hal\`asz and J.J.M. Verbaarschot, Phys. Rev. Lett. 
{\bf 74} (1995) 3920-3923 [hep-lat/9501025].

\bibitem{GMMW} T. Guhr, J.-Z. Ma, S. Meyer and T. Wilke, Phys. Rev. {\bf 
    D59} (1999) 054501 [hep-lat/9806003]. 

\bibitem{DHKM} P.H. Damgaard, U.M. Heller, A. Krasnitz and T. Madsen,
Phys. Lett. {\bf B440} (1998) 129-135 [hep-lat/9803012].

\bibitem{Jack}B. Vanderheyden and A.D. Jackson,
  Phys. Rev. {\bf D62} (2000) 094010 [hep-ph/0003150];\newline 
S. Pepin and  A. Sch\"afer, Eur. Phys. J. {\bf A10} (2001) 
  303-308 [hep-ph/0010225].

\bibitem{DNII} P.H. Damgaard and S. Nishigaki,
Phys. Rev. {\bf D57} (1998) 5299-5302 [hep-th/9711096].

\bibitem{ADMN}G. Akemann, P.H. Damgaard, U. Magnea and S. Nishigaki,
Nucl. Phys. {\bf B487} (1997) 721-738 [hep-th/9609174].

\bibitem{AD}  G. Akemann and P.H. Damgaard, Nucl. 
Phys. {\bf B528} (1998) 411-431 [hep-th/9801133];\newline
J. Christiansen, Nucl. Phys. {\bf B547} (1999) 329-342 [hep-th/9809194]. 

\bibitem{Szabo}R. Szabo, Nucl. Phys. {\bf B598} 
(2001) 309-347 [hep-th/0009237].

\bibitem{ADDV}G. Akemann, P.H. Damgaard, D. Dalmazi and
  J.J.M. Verbaarschot,  Nucl. Phys. {\bf B601} (2001) 77-124 [hep-th/0011072].

\bibitem{DV}D. Dalmazi and J.J.M. Verbaarschot, hep-th/0101035.

\bibitem{AV} G. Akemann and G. Vernizzi,  Nucl. Phys. {\bf B583}
  (2000) 739-757 [hep-th/0002148].

\bibitem{FKS98}Y.V. Fyodorov, B.A. Khoruzhenko and H.-J. Sommers,
Ann. Inst. Henri Poincar\'e 
{\bf 68} (1998) 449-489 [chao-dyn/9802025].

\bibitem{FGIL}P. Di Francesco, M. Gaudin, C. Itzykson and F. Lesage,
  Int. J. Mod. Phys. {\bf A9} (1994) 4257-4351.

\bibitem{Mehta}M.L. Mehta, {\it Random Matrices}, Academic Press,
  London 1991.

\bibitem{Sze}G. Szeg\"o, {\it Orthogonal Polynomials}, American
  Mathematical Society, Providence RI 1939.

\bibitem{AK}G. Akemann and E. Kanzieper,  Phys. Rev. Lett. {\bf 85}
  (2000) 1174-1177 [hep-th/0001188].

\bibitem{FKS96}Y.V. Fyodorov, B.A. Khoruzhenko and H.-J. Sommers,
Phys. Lett. {\bf A226} (1997) 46-52 [cond-mat/9606173].

\bibitem{ADII} G. Akemann and P.H. Damgaard,  
Phys. Lett. {\bf B432} (1998) 390-396 [hep-th/9802174].


\bibitem{BZ} E. Br\'ezin and A. Zee, Nucl. Phys. {\bf B402} (1993) 613-627.

\bibitem{AJM} J. Ambj\o rn, J. Jurkiewicz and Yu. Makeenko,
            Phys. Lett. {\bf B251} (1990) 517-527.

\bibitem{Shu}P. Shukla, cond-mat/0105007.



\end{thebibliography}
\end{document}